\documentclass[structabstract]{aa}
\usepackage{graphicx}
\usepackage{txfonts}
\usepackage{longtable}
\begin{document}
   \title{Thin accretion disks around neutron and quark stars}


   \author{Z. Kov\'{a}cs
          \inst{1,}
          \inst{2}
          \and
          K. S. Cheng\inst{3}
          \and
          T. Harko\inst{3}}

   \institute{Max-Planck-Institut f\"{u}r Radioastronomie, Auf dem H\"{u}gel 69,
53121 Bonn, Germany\\
              \email{zkovacs@mpifr-bonn.mpg.de}
              \and
              Department of Experimental Physics, University of Szeged, D\'{o}m T%
\'{e}r 9, Szeged 6720, Hungary
         \and
             Department of Physics and Center for Theoretical and Computational Physics, The University of Hong Kong, Hong Kong, P. R. China\\
             \email{hrspksc@hkucc.hku.hk}
             \email{harko@hkucc.hku.hk}
             }

\date{\today}


  \abstract
{The possibility of observationally discriminating between various types of neutron stars, described by different equations of state of the nuclear matter, as well as differentiating neutron stars from other types of
exotic objects, like, for example, quark stars, is one of the fundamental problems in contemporary astrophysics.}
{We consider and investigate carefully  the possibility that different types of rapidly rotating neutron stars,
as well as other type of compact general relativistic objects, can be differentiated from the study of
the emission properties of the accretion disks around them.}
{We obtain the energy flux, the temperature distribution and the
emission spectrum from the accretion disks around several classes of rapidly rotating neutron stars, described by different equations of state of the neutron matter, and for quark stars,
described by the MIT bag model equation of state and in the CFL (Color-Flavor-Locked) phase, respectively.}
{Particular signatures appear in the electromagnetic spectrum, thus leading to the possibility of
directly testing the equation of state of the dense matter by using astrophysical
observations of the emission spectra from accretion disks.}
   {}


\maketitle
%

\section{Introduction}

The quark structure of the nucleon indicates the possibility of a phase transition of confined hadronic matter to absolutely stable strange quark matter at high densities (\cite{It70, Bo71, Wi84}), which is referred to as strange matter hypothesis. If the hypothesis is true, then some neutron stars should actually be stars made of strange quark matter (strange stars) (\cite{Al86,Ha86}). For a general review of strange star
properties see Cheng et al. (1998a).

There are several proposed mechanisms for the formation of quark stars.
Quark stars are expected to form during the collapse of the core of a
massive star, after the supernova explosion, as a result of a first or second
order phase transition, resulting in deconfined quark matter (\cite{Da}). The
proto-neutron star core or the neutron star core is a favorable environment
for the conversion of ordinary matter to strange quark matter (\cite{ChDa,Cha09}).
Another possibility is that some neutron stars in low-mass X-ray binaries
can accrete sufficient mass to undergo a phase transition to become strange
stars (\cite{Ch96}). This mechanism has also been proposed as a source of
radiation emission for cosmological $\gamma $-ray bursts (\cite{Ch98a}).

Based on numerical integration
of the general relativistic hydrostatic equilibrium equations a
complete description of the basic astrophysical properties (mass,
radius, eccentricity, Keplerian frequency etc.) of both static and
rotating strange stars has been obtained, for different values of
the bag constant and for different equations of state of the
strange star (\cite{Wi84, Ha86, Go00, De98, HaCh02}). Rotational properties can discriminate between
neutron and quark stars. Strange stars can reach much shorter periods than
neutron stars, of the order of $0.5$ ms (\cite{Ch98}). $r$-mode instabilities in rapidly rotating strange stars lead to specific signatures in the evolution of pulsars with periods below $2.5$ ms, and some data on pulsar properties are consistent with this assumption (\cite{Ma00}).
Strange stars could have a radius significantly less than that of neutron stars (\cite{Ch98}).

Photon emissivity is the basic parameter for determining macroscopic
properties of stellar type objects.  Because of the very high plasma frequency $\omega _p$ near the strange
matter edge, photon emissivity of strange matter is very low (\cite{Al86}).
The spectrum of equilibrium photons is very hard, with $%
\hbar \omega >20$ MeV. The problem of the soft photon emissivity of quark matter at the surface
of strange stars has also been considered (\cite{ChHa03, HaCh05}). By taking into account the Landau-Pomeranchuk-Migdal effect and the absorption of the radiation in the
external electron layer, the emissivity of the quark matter can be six
orders of magnitude lower than the equilibrium black body radiation.

The Coulomb barrier at the quark surface of a hot strange star may also be a
powerful source of $e^{+}e^{-}$ pairs, which are created in the extremely
strong electric field of the barrier. At surface temperatures of around $%
10^{11}$ K, the luminosity of the outflowing plasma may be of the order $%
\sim 10^{51}$ ergs$^{-1}$ (\cite{Us98a, Us98b,HaCh06}). Moreover,
for about one day for normal quark matter and for up to a
hundred years for superconducting quark matter, the thermal luminosity from
the star's surface, due to both photon emission and $e^{+}e^{-}$ pair
production, may be orders of magnitude higher than the Eddington limit (\cite{PaUs02}).

However, despite these very specific signatures for quark stars, a definite method for discriminating them with respect to the neutron stars is still missing.

It is generally expected that most of the astrophysical objects grow
substantially in mass via accretion. Accretion discs are flattened astronomical objects made of rapidly rotating gas which slowly spirals onto a central gravitating body, with its
gravitational energy degraded to heat. A fraction of the heat converts into
radiation, which partially escapes, and cools down the accretion disc.
The only information that we have about accretion disk physics comes from
this radiation, when it reaches radio, optical and $X$-ray telescopes,
allowing astronomers to analyze its electromagnetic spectrum, and its time
variability. The efficient cooling via the radiation over the disk surface prevents the disk
from cumulating the heat generated by stresses and dynamical
friction. In turn, this equilibrium causes the disk to stabilize
its thin vertical size. The thin disk has an inner edge at the
marginally stable orbit of the compact object potential, and the
accreting plasma has a Keplerian motion in higher orbits (\cite{PaTh74, Th74}).

Recent astrophysical observations have shown that around
many compact objects, as well as around black hole candidates, there
are gas clouds surrounding the central compact object, and an associated
accretion disc, on a variety of scales that could range up to a tenth of a parsec or even to a few
hundred parsecs (\cite{UrPa95}). These clouds are assumed to form a
geometrically and optically thick torus (or warped disc), which absorbs most
of the ultraviolet radiation and the soft X-rays. The gas exists in either
the molecular or the atomic phase.  Hence,
important astrophysical information can be obtained from the observation of
the motion of the gas streams in the gravitational field of compact objects.


The first comprehensive theory of accretion disks was
constructed in Shakura \& Sunyaev (1973) by using a Newtonian approach. This theory was extended to the general
relativistic models of the mass accretion onto rotating black holes (\cite%
{NoTh73}). These pioneering works developed thin steady-state accretion
disks, where the accreting matter moves in Keplerian orbits. The
hydrodynamical equilibrium in the disk is maintained by an efficient cooling
mechanism via radiation transport. The photon flux emitted by the disk
surface was studied under the assumption that the disk emits a black body
radiation. The properties of radiant energy flux over the thin accretion
disks were further analyzed,  with the
effects of the photon capture by the hole on the spin evolution
taken into account as well (\cite{PaTh74, Th74}). In these works the efficiency with which black holes
convert rest mass into outgoing radiation in the accretion process was also
computed.

The disc and boundary layer luminosity for accreting rapidly rotating neutron stars with low magnetic fields in a fully general relativistic manner were considered in \cite{ThDa98}. The effect of the quadrupole component in the mass distribution of a rapidly rotating neutron star on the energy release in the equatorial (or boundary) layer on the surface of the accreting star and in the accretion disk in the cases where the stellar radius is smaller (or larger) than the radius of the last stable circular orbit was analyzed in \cite{SiSu98}. The temperature profiles of (thin) accretion disks around rapidly rotating neutron stars (with low surface magnetic fields), taking into account the full effects of general relativity were obtained in \cite{Baetal00}. The calculations suggest that the neutron star in Cygnus X-2 rotates close to the centrifugal mass-shed limit. General relativistic spectra from accretion disks around rotating neutron stars in the appropriate spacetime geometry for several different equations of state, spin rates, and masses of the compact object have been also computed (\cite{BMT01}). The spectra from the accretion disks around a rapidly rotating neutron star, by taking into account the Doppler shift, the gravitational redshift and the light-bending effects  were obtained in \cite{Bhaetal01}. Light bending significantly modifies the high-energy part of the spectrum. These results could be important for modeling the observed X-ray spectra of low-mass X-ray binaries, containing fast-spinning neutron stars. The temperature profiles of accretion discs around rapidly rotating strange stars, using constant gravitational mass equilibrium sequences of these objects, and considering the full effect of general relativity were computed in \cite{Bom}. The structure parameters and temperature profiles obtained were compared with those of neutron stars, as an attempt to provide signatures for distinguishing between the two. The obtained results implies the possibility of distinguishing these objects from each other by sensitive observations through future X-ray detectors.  The accretion disc temperature profiles, disc luminosity and boundary layer luminosity for rapidly rotating neutron stars, considering the full effect of general relativity, were calculated in \cite{Bh02}.

The emissivity properties of the accretion disks have been recently investigated for a large class of compact objects, such as rotating and non-rotating boson or fermion stars (\cite{To02,YuNaRe04,Gu06}), as well as for the modified $f(R)$ type theories of gravity (\cite{mod}), for brane world black holes (\cite{mod1}), and for wormholes (\cite{mod2, mod3}). The radiation power per unit area, the temperature of the disk and the spectrum of the emitted radiation were given, and compared with the case of a Schwarzschild or Kerr-Newman black holes of an equal mass.

It is the purpose of the present paper to consider a comparative systematic study of the properties of the thin accretion disks around rapidly rotating neutron and strange stars, respectively, and to obtain the basic physical parameters describing the disk, like the emitted energy flux, the temperature distribution on the surface of the disk, as well as the spectrum of the emitted equilibrium black body radiation.

In order to obtain the emissivity properties of the disk, the metric outside the rotating general relativistic stars must be determined. In the present study we calculate the equilibrium configurations of the rotating neutron and quark stars by using the RNS code, as introduced in \cite{SteFr95}, and discussed in detail in \cite{Sterev}. This code was used for the study of different models of rotating neutron stars in \cite{No98} and for the study of the rapidly rotating strange stars (\cite{Ste99}). The software provides the metric potentials for various types of compact rotating general relativistic objects, which can be used to obtain the physical properties of the accretion disks. Particular signatures appear in the electromagnetic spectrum, thus leading to the possibility of
directly testing the equation of state of the dense matter by using astrophysical
observations of the emission spectra from accretion disks.

The present paper is organized as follows.  The properties of the general relativistic thin accretion disks onto compact objects are briefly described in Section II. The equations of state used in the present study are presented in Section III. In Section IV we consider the radiation
flux, spectrum and efficiency of thin accretion disks onto several classes
of neutron stars and quark stars. We discuss and conclude our results in Section V.

\section{Thin accretion disks onto general relativistic compact objects}


A thin accretion disk is an accretion disk such that in
cylindrical coordinates $(r,\phi ,z)$ most of the
matter lies close to the radial plane.
For the thin accretion disk its vertical size (defined along the $z$-axis)
is negligible, as compared to its horizontal extension (defined along the
radial direction $r$), i.e, the disk height $H$, equal to the maximum
half thickness of the disk, is always much smaller than the characteristic radius $R$ of
the disk, $H \ll R$ (\cite{NoTh73}). The thin disk is in hydrodynamical equilibrium, and the pressure gradient
and a vertical entropy gradient in the accreting matter are negligible.
In the steady-state accretion disk models, the mass accretion rate $\dot{M%
}_{0} $ is supposed to be constant in time, and the physical quantities of
the accreting matter are averaged over a characteristic time scale, e.g. $%
\Delta t$, and over the azimuthal angle $\Delta \phi =2\pi $, for a total period of
the orbits and for the height $H$. The plasma moves in Keplerian orbits around
the compact object, with a rotational velocity $\Omega $. The plasma
particles have a specific energy $\widetilde{E}$, and a specific angular
momentum $\widetilde{L}$, which depend only on the radii of the orbits. The
four-velocity of the particles is $u^{\mu }$, and the disk has an
averaged surface density $\Sigma $. The accreted matter is modeled by an
anisotropic fluid source, where the density $\rho _{0}$ (the specific heat
is neglected), the energy flow vector $q^{\mu }$, and the stress tensor $%
t^{\mu \nu }$ are measured in the averaged rest-frame. The energy-momentum
tensor takes the form (\cite{NoTh73})
\begin{equation}
T^{\mu \nu }=\rho _{0}u^{\mu }u^{\nu }+2u^{(\mu }q^{\nu )}+t^{\mu \nu }\;,
\end{equation}
where $u_{\mu }q^{\mu }=0$, $u_{\mu }t^{\mu \nu }=0$. The four-vectors of
the energy and of the angular momentum flux are defined by
$-E^{\mu }\equiv T_{{}}^{\mu }{}_{\nu }(\partial /\partial t)^{\nu }$ and $
J^{\mu }\equiv T_{{}}^{\mu }{}_{\nu }(\partial /\partial \phi )^{\nu }$,
respectively. The four dimensional conservation laws
of the rest mass, of the energy and of the angular momentum of the plasma
provide the structure equations of the thin disk. By integrating the
equation of the rest mass conservation, $\nabla _{\mu }(\rho _{0}u^{\mu })=0$, %
 it follows that the time averaged
 accretion rate  $\dot{M_{0}}$ is independent of the disk radius:
\begin{equation}
\dot{M_{0}}\equiv -2\pi r\Sigma u^{r}=\mbox{const}\;,  \label{conslawofM}
\end{equation}%
where a dot represents the derivative with respect to the time coordinate (\cite{PaTh74}).
The averaged rest mass density  is defined by
$\Sigma (r)=\int_{-H}^{H}\langle \rho _{0}\rangle dz$,
where $\langle \rho _{0}\rangle $ is the rest mass density averaged  over $\Delta
t$ and $2\pi $. The conservation law $\nabla _{\mu }E^{\mu }=0$ of the
energy can be written in an integral form as
\begin{equation}
\lbrack \dot{M}_{0}\widetilde{E}-2\pi r\Omega W_{\phi }{}^{r}]_{,r}=4\pi
\sqrt{-g}F\widetilde{E}\;\;,  \label{conslawofE}
\end{equation}%
where a comma denotes the derivative with respect to the radial coordinate $r$. Eq.~(\ref{conslawofE}) shows the balance between the energy transported by the rest mass flow, the dynamical stresses in the disk, and the energy radiated away
from the surface of the disk, respectively. The torque $W_{\phi }{}^{r}$ in Eq.~(\ref{conslawofE}) is given by
\begin{equation}
W_{\phi }{}^{r}=\int_{-H}^{H}\langle t_{\phi }{}^{r}\rangle dz,
\end{equation}%
where $\langle t_{\phi }{}^{r}\rangle $ is the $\phi -r$ component of the stress
tensor, averaged over $\Delta t$ and over a $2\pi $ angle. The law of the angular momentum
conservation, $\nabla _{\mu }J^{\mu }=0$,  states in its integral form the
balance of the three forms of the angular momentum transport,
\begin{equation}
\lbrack \dot{M}_{0}\widetilde{L}-2\pi rW_{\phi }{}^{r}]_{,r}=4\pi \sqrt{-g}F%
\widetilde{L}\;\;.  \label{conslawofL}
\end{equation}

By eliminating $W_{\phi}{}^{r}$ from Eqs. (\ref{conslawofE}) and (\ref%
{conslawofL}), and by applying the universal energy-angular momentum relation
$dE=\Omega dJ$ for circular geodesic orbits in the form $\widetilde{E}%
_{,r}=\Omega\widetilde{L}_{,r}$, the flux of the
radiant energy over the disk can be expressed in terms of the specific energy, angular
momentum and the angular velocity of the black hole. Then the flux integral
leads to the expression of the energy flux $F(r)$, which is given by (\cite{PaTh74, Th74})
\begin{equation}
F(r)=-\frac{\dot{M}_0}{4\pi\sqrt{-g}} \frac{\Omega_{,r}}{(\widetilde{E}%
-\Omega\widetilde{L})^{2}} \int_{r_{ms}}^{r}(\widetilde{E}-\Omega\widetilde{L%
})\widetilde{L}_{,r}dr\;,  \label{F}
\end{equation}
where the no-torque inner boundary conditions were also prescribed (\cite{PaTh74}). This means that the torque vanishes at the inner edge of the disk,
since the matter at the marginally stable orbit $r_{ms}$ falls
freely into the black hole, and cannot exert considerable torque on the
disk. The latter assumption is valid as long as strong magnetic fields do
not exist in the plunging region, where matter falls into the hole.

Once the geometry of the space-time is known, we can derive the time
averaged radial distribution of photon emission for accretion disks around
black holes, and determine the efficiency of conversion of the rest mass into
outgoing radiation. After obtaining the radial dependence of
the angular velocity $\Omega $, of the specific energy $\widetilde{E}$ and
of the specific angular momentum $\widetilde{L}$ of the particles moving on
circular orbits around the black holes, respectively, we can compute the
flux integral given by Eq.~(\ref{F}).

Let us consider an arbitrary stationary and axially symmetric geometry,
\begin{equation}
ds^{2}=g_{tt}dt^{2}+g_{t\phi }dtd\phi +g_{rr}dr^{2}+g_{\theta \theta
}d\theta ^{2}+g_{\phi \phi }d\phi ^{2}\;,  \label{ds2rcoappr}
\end{equation}%
where in the equatorial approximation ($|\theta -\pi /2|\ll 1$)the metric functions $g_{tt}$, $g_{t\phi }$, $g_{rr}$, $g_{\theta
\theta }$ and $g_{\phi \phi }$ depend only on the radial coordinate $r$. The geodesic equations take the form
\begin{equation}
 \frac{dt}{d\tau }=\frac{\widetilde{E}g_{\phi \phi }+%
\widetilde{L}g_{t\phi }}{g_{t\phi }^{2}-g_{tt}g_{\phi \phi }},
 \frac{d\phi }{d\tau }=-\frac{\widetilde{E}g_{t\phi }+%
\widetilde{L}g_{tt}}{g_{t\phi }^{2}-g_{tt}g_{\phi \phi }},
\end{equation}%
and
\begin{equation}
g_{rr}\left( \frac{dr}{d\tau }\right) ^{2}=V(r),
\end{equation}%
respectively, where $\tau $ is the affine parameter, and the potential term $V(r)$ is defined by
\begin{equation}
V(r)\equiv -1+\frac{\widetilde{E}^{2}g_{\phi \phi }+2\widetilde{E}%
\widetilde{L}g_{t\phi }+\widetilde{L}^{2}g_{tt\texttt{}}}{g_{t\phi
}^{2}-g_{tt}g_{\phi \phi }}\;.
\end{equation}

For circular orbits in the equatorial plane the conditions $V(r)=0$ and $V_{,r}(r)=0$, respectively, must
hold for all disk configurations. From these conditions one can obtain the specific energy $\widetilde{E}$, the specific angular
momentum $\widetilde{L}$ and the angular velocity $\Omega $ of the particles
moving on circular orbits around spinning general relativistic stars as
\begin{eqnarray}
\widetilde{E} &=&-\frac{g_{tt}+g_{t\phi }\Omega }{\sqrt{-g_{tt}-2g_{t\phi
}\Omega -g_{\phi \phi }\Omega ^{2}}}\;,  \label{tildeE} \\
\widetilde{L} &=&\frac{g_{t\phi }+g_{\phi \phi }\Omega }{\sqrt{%
-g_{tt}-2g_{t\phi }\Omega -g_{\phi \phi }\Omega ^{2}}},  \label{tildeL} \\
\Omega  &=&\frac{d\phi }{dt}=\frac{-g_{t\phi ,r}+\sqrt{(g_{t\phi
,r})^{2}-g_{tt,r}g_{\phi \phi ,r}}}{g_{\phi \phi ,r}}\;.\label{Omega}
\end{eqnarray}%
The marginally stable orbits $r_{ms}$ around the central object are determined by
the condition
$\left.V_{,rr}(r)\right|_{r=r_{ms}}=0$, which is equivalent to the equation%
\begin{equation}
\left[\widetilde{E}^{2}g_{\phi \phi ,rr}+2\widetilde{E}\widetilde{L}g_{t\phi ,rr}+%
\widetilde{L}^{2}g_{tt ,rr}-(g_{t\phi }^{2}-g_{tt}g_{\phi \phi
})_{,rr}\right]_{r=r_{ms}}=0\;.  \label{stable}
\end{equation}

By inserting Eqs.~(\ref{tildeE})-(\ref{tildeL}) into Eq.~(\ref{stable}), and
solving the resulting equation for $r_{ms}$, we obtain the marginally stable
orbits, once the metric coefficients $g_{tt}$, $g_{t\phi }$ and $%
g_{\phi \phi }$ are explicitly given.



The accreting matter in the steady-state thin disk model is supposed to be
in thermodynamical equilibrium. Therefore the radiation emitted by the disk
surface can be considered as a perfect black body radiation, where the
energy flux is given by $F(r)=\sigma T^{4}(r)$ ($\sigma $ is the
Stefan-Boltzmann constant), and the observed luminosity $L\left( \nu \right) $ has a redshifted black body spectrum (\cite{To02}):
\begin{equation}
L\left( \nu \right) =4\pi d^{2}I\left( \nu \right) =\frac{8}{\pi c^2 }\cos \gamma \int_{r_{i}}^{r_{f}}\int_0^{2\pi}\frac{\nu^{3}_e r d\phi dr }{\exp \left( h\nu_e/T\right) -1}.
\end{equation}

Here $d$ is the distance to the source, $I(\nu )$ is the Planck
distribution function, $\gamma $ is the disk inclination angle, and $r_{i}$
and $r_{f}$ indicate the position of the inner and outer edge of the disk,
respectively. We take $r_{i}=r_{ms}$ and $r_{f}\rightarrow \infty $, since
we expect the flux over the disk surface vanishes at $r\rightarrow \infty $
for any kind of general relativistic compact object geometry. The emitted frequency is given by $\nu_e=\nu(1+z)$, where the redshift factor can be written as
\begin{equation}
1+z=\frac{1+\Omega r \sin \phi \sin \gamma }{\sqrt{ -g_{tt} - 2 \Omega g_{t\phi} - \Omega^2 g_{\phi\phi}}}
\end{equation}
where we have neglected the light bending (\cite{Lu79,BMT01}).

The flux and the emission spectrum of the accretion disks around compact objects
satisfy some simple scaling relations, with respect to the simple scaling
transformation of the radial coordinate, given by $r\rightarrow \widetilde{r}=r/M$,
where $M$ is the mass of the black hole. Generally, the metric tensor
coefficients are invariant with respect of this transformation, while the
specific energy, the angular momentum and the angular velocity transform as $%
\widetilde{E}\rightarrow \widetilde{E}$, $\widetilde{L}\rightarrow M\widetilde{L}$ and $%
\Omega \rightarrow \widetilde{\Omega}/M$, respectively. The flux scales as $F(r)\rightarrow F(%
\widetilde{r})/M^{4}$, giving the simple transformation law of the temperature as $%
T(r)\rightarrow T\left( \widetilde{r}\right) /M$. By also rescaling the frequency
of the emitted radiation as  $\nu \rightarrow \widetilde{\nu}=\nu /M$,
the luminosity of the disk is given by $L\left( \nu \right) \rightarrow
L\left( \widetilde{\nu}\right) /M$. On the other hand, the flux is proportional
to the accretion rate $\dot{M}_{0}$, and therefore an increase in the
accretion rate leads to a linear increase in the radiation emission flux
from the disc.

The efficiency $\epsilon $ with which the central object converts rest mass into outgoing radiation is the other important physical parameter characterizing the properties of the accretion disks. The efficiency
is defined by the ratio of two rates measured at infinity: the rate of the
radiation of the energy of the photons escaping from the disk surface to infinity, and the rate at which mass-energy is transported to the compact object. If all the emitted photons can escape to infinity, the efficiency depends only on the specific energy measured at the marginally stable orbit $r_{ms}$,
\begin{equation}
\epsilon = 1 - \left.\widetilde{E}\right|_{r=r_{ms}}\;.  \label{epsilon}
\end{equation}

There are some solutions for the geometry of the neutron and quark stars where  $r_{ms}$ is at the surface of the star or even takes values less than the surface radius of the central object. In this case the inner edge of the disk touches the surface of the star and plasma under the effect of any perturbation due to hydro- or magnetohydrodynamic instabilities in the disk will leave the disk and hit the surface. Then the energy $\widetilde{E}_{e}$ transferred to the star from the disk is measured at the radius $R_e$ of the star, and the efficiency takes the form
\begin{equation}
\epsilon = 1 - \widetilde{E}_{e}\;,  \label{epsilon2}
\end{equation}
where $\widetilde{E}_{e}=\left.\widetilde{E}\right|_{r=R_e}$. For Schwarzschild black holes the efficiency is about 6\%, no matter if we
consider the photon capture by the black hole, or not. Ignoring the capture
of radiation by the black hole, $\epsilon$ is found to be 42\% for rapidly
rotating black holes, whereas the efficiency is 40\% with photon capture in
the Kerr potential. For neutron and quark stars the efficiency is varying in a broader range, especially if we take into account that $\widetilde{E}_{ms}$ and $\widetilde{E}_{e}$ can have very different values for different neutron and quark star models.

\section{Equations of state}\label{eos}

In order to obtain a consistent and realistic physical description of the rotating general relativistic neutron and quark stars, as a first step we have to adopt the equations of state for the dense neutron and quark matter, respectively. In the present study of the accretion disks onto rapidly rotating neutron and quark stars we consider the following equations of state:

1) Akmal-Pandharipande-Ravenhall  (APR) EOS (\cite{Ak98}). EOS APR has been obtained by using the variational chain summation methods and the  Argonne $v_{18}$ two-nucleon interaction. Boost corrections to the two-nucleon interaction, which give the leading relativistic effect of order $(v/c)^2$, as well as three-nucleon interactions, are also included in the nuclear Hamiltonian.  The density range is from $2\times 10^{14}$ g/cm$^3$ to $2.6\times 10^{15}$ g/cm$^3$. The maximum mass limit in the static case for this EOS is $2.20 M_{\odot}$.  We join this equation of state to the composite BBP ($\epsilon /c^2>4.3\times10^{11}$g/cm$^3$) (\cite{Ba71a}) - BPS ($10^4$ g/cm$^3$ $<4.3\times 10^{11}$g/cm$^3$) (\cite{Ba71b}) - FMT ($\epsilon/c^2<10^4$ g/cm$^3$) (\cite{Fe49}) equations of state, respectively.

2) Douchin-Haensel (DH) EOS (\cite{DoHa01}). EOS DH is an equation of state of the neutron star matter, describing both the neutron star crust and the liquid core. It is based on the effective nuclear interaction SLy of the Skyrme type, which is particularly suitable for the application to the calculation of the properties of very neutron rich matter. The structure of the crust, and its EOS, is calculated in the zero temperature approximation, and under the assumption of the ground state composition. The EOS of the liquid core is calculated assuming (minimal) $npe\mu $ composition. The density range is from $3.49\times 10^{11}$ g/cm$^3$ to $4.04\times 10^{15}$ g/cm$^3$. The minimum and maximum masses of the static neutron stars for this EOS are $0.094M_{\odot}$ and $2.05 M_{\odot}$, respectively.

3) 	Shen-Toki-Oyamatsu-Sumiyoshi (STOS) EOS (\cite{Shen}). The STOS equation of state of nuclear matter is obtained by using the relativistic mean field theory with nonlinear $\sigma $ and $\omega $ terms in a wide density and temperature range, with various proton fractions. The EOS was specifically designed for the use of supernova simulation and for the neutron star calculations. The  Thomas-Fermi approximation is used to describe inhomogeneous matter, where heavy nuclei are formed together with free nucleon gas. We consider the STOS EOS for several temperatures, namely $T=0$, $T=0.5$ and $T=1$ MeV, respectively. The temperature is mentioned for each STOS equation of state, so that, for example, STOS 0 represents the STOS EOS for $T=0$. For the proton fraction we chose  the value $Y_p=0.001$ in order to avoid the negative pressure regime for low baryon mass densities.

4) Relativistic Mean Field (RMF) equations of state with isovector scalar mean field corresponding to the $\delta $-meson- RMF soft and RMF stiff EOS (\cite{Kubis}).  While the $\delta $-meson mean field vanishes in symmetric nuclear matter, it can influence properties of asymmetric nuclear matter in neutron stars. The Relativistic mean field contribution due to the $\delta $-field to the nuclear symmetry energy is negative. The energy per particle of neutron matter is then larger at high densities than the one with no $\delta $-field included. Also, the proton fraction of $\beta $-stable matter increases. Splitting of proton and neutron effective masses due to the $\delta $-field can affect transport properties of neutron star matter. The equations of state can be parameterized by the coupling parameters $C_{\sigma }^2=g_{\sigma }^2/m_{\sigma }^2$, $C_{\omega }^2=g_{\omega }^2/m_{\omega }^2$, $\bar{b}=b/g_{\sigma}^3$ and $\bar{c}=c/g_{\sigma}^4$, where $m_{\sigma }$ and $m_{\omega }$ are the masses of the respective mesons, and $b$ and $c$ are the coefficients in the potential energy $U\left(\sigma \right)$ of the $\sigma $-field. The soft RMF EOS is parameterized by $C_{\sigma }^2=1.582$ fm$^2$, $C_{\omega }^2=1.019$ fm$^2$, $\bar{b}=-0.7188$ and $\bar{c}=6.563$, while the stiff RMF EOS is parameterized by $C_{\sigma }^2=11.25$ fm$^2$, $C_{\omega }^2=6.483$ fm$^2$, $\bar{b}=0.003825$ and $\bar{c}=3.5\times 10^{-6}$, respectively.

5) Baldo-Bombaci-Burgio (BBB) EOS (\cite{baldo}). The BBB EOS is an EOS for asymmetric nuclear matter, derived from the Brueckner-Bethe-Goldstone many-body theory with explicit three-body forces. Two EOS's are obtained, one corresponding to the Argonne AV14 (BBBAV14), and the other to the Paris two-body nuclear force (BBBParis), implemented by the Urbana model for the three-body force. The maximum static mass configurations are $M_{max} = 1.8 M_{\odot}$ and  $M_{max} = 1.94 M_{\odot}$ when the AV14 and Paris interactions are used, respectively.  The onset of direct Urca processes occurs at densities $n\geq 0.65$ fm$^{-3}$ for the AV14 potential and $n\geq 0.54$ fm$^{-3}$ for the Paris potential. The comparison with other microscopic models for the EOS shows noticeable differences. The density range is from $1.35\times 10^{14}$ g/cm$^3$ to $3.507\times 10^{15}$ g/cm$^3$.

6) Bag model equation of state (Q) EOS (\cite{Wi84,Ch98}). For the description of the quark matter we adopt first a simple phenomenological description, based on the MIT bag model equation of state, in which the pressure $p$ is related to the energy density $\rho $ by
\begin{equation}
p=\frac{1}{3}\left(\rho-4B\right),
\end{equation}
where $B$ is the difference between the energy density of the perturbative and
non-perturbative QCD vacuum (the bag constant), with the value $B=4.2\times 10^{14}$ g/cm$^3$.

7) Color-Flavor-Locked (CFL) EOS It is generally agreed today that the color-flavor-locked (CFL) state is
likely to be the ground state of matter, at least for asymptotic densities,
and even if the quark masses are unequal (\cite{cfl1,cfl2,cfl3}). Moreover,
the equal number of flavors is enforced by symmetry, and electrons are
absent, since the mixture is automatically neutral. By assuming that the
mass $m_{s}$ of the $s$ quark is not large as compared to the chemical
potential $\mu $, the thermodynamical potential of the quark matter in CFL
phase can be approximated as (\cite{LuHo02})
\begin{equation}
\Omega _{CFL}=-\frac{3\mu ^{4}}{4\pi ^{2}}+\frac{3m_{s}^{2}}{4\pi ^{2}}-%
\frac{1-12\ln \left( m_{s}/2\mu \right) }{32\pi ^{2}}m_{s}^{4}-\frac{3}{\pi
^{2}}\Delta ^{2}\mu ^{2}+B,
\end{equation}%
where $\Delta $ is the gap energy. With the use of this expression the
pressure $P$ of the quark matter in the CFL phase can be obtained as an
explicit function of the energy density $\varepsilon $ in the form (\cite{LuHo02})
\begin{equation}
P=\frac{1}{3}\left( \varepsilon -4B\right) -\frac{2\Delta ^{2}\delta ^{2}}{\pi
^{2}}+\frac{m_{s}^{2}\delta ^{2}}{2\pi ^{2}},
\end{equation}
where
\begin{equation}
\delta ^{2}=-\alpha +\sqrt{\alpha ^{2}+\frac{4}{9}\pi ^{2}\left( \varepsilon
-B\right) },
\end{equation}%
and $\alpha =-m_{s}^{2}/6+2\Delta ^{2}/3$. In the following the value of the gap energy $\Delta $ considered in each case will be also mentioned for the CFL equation of state, so that, for example, CFL150 represents the CFL EOS with $\Delta =150$ MeV. For the mass of the strange quark we adopt the value $m_s=150$ MeV. The maximum mass of the stars in the CFL model is given by $M_{\max}=1.96M_{\odot}\left(1+\delta \right)/\sqrt{B_{60}}$, with $\delta =0.15\left(\Delta/100\; {\rm MeV}\right)^2\left(60\;{\rm MeV}\;{\rm fm}^{-3}/B\right)$ (\cite{HoLu04}). Hence large maximum masses ($M>4M_{\odot}$) can be found for standard values of $B$ and $\Delta \geq 250$ MeV. For a discussion of the maximum mass of both rotating and non-rotating general relativistic objects see \cite{Sterev}. For rotating stars the maximum mass can be of the order of $M_{max}\approx 6.1M_{\odot}\left(2\times 10^{14}\;{\rm g}\;{\rm cm}^{-3}/\rho _m\right)^{1/2}$, where $\rho _m$ is the matching density.

The pressure-density relation is presented for the considered equations of state in Fig.~{\ref{fig1}.

\begin{figure}[tbp]
\centering
\includegraphics[width=8.15cm]{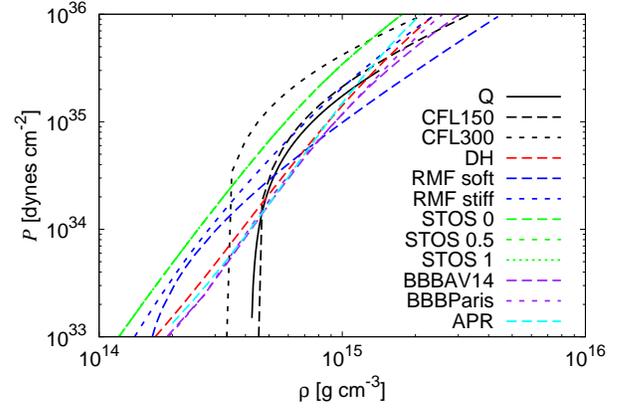}
\caption{Pressure as a function of density (in a logarithmic scale) for the equations of state  Q, CFL150, CFL300, DH, RMF soft, RMF stiff, STOS 0. STOS 0.5, STOS 1, BBBAV14, BBBParis and APR, respectively.}
\label{fig1}
\end{figure}

\section{Thin disk accretion onto neutron and quark stars }

In the present Section we consider the properties of the general relativistic thin accretion disks formed around neutron stars and quark stars. The metric outside the rotating compact general relativistic stars can be described, in quasi-isotropic coordinates, as
\begin{equation}
ds^2=-e^{\bar{\gamma }+\bar{\rho }}dt^2+e^{2\bar{\alpha } }\left(d\bar{r}^2+\bar{r}^2d\theta ^2\right)+e^{\bar{\gamma }-\bar{\rho }}\bar{r}^2\sin ^2\theta \left(d\phi -\bar{\omega}dt\right)^2,
\end{equation}
where the metric potentials $\bar{\gamma }$, $\bar{\rho }$, $\bar{\alpha }$ and the angular velocity of the stellar fluid relative to the local inertial frame  $\bar{\omega}$ are all functions of the quasi-isotropic radial coordinate $\bar{r}$ and of the polar angle $\theta $. The RNS code produces the metric functions in a quasi-spheroidal coordinate system, as functions of the parameter $s=\bar{r}/\left(\bar{r}+\bar{r}_e\right)$, where $\bar{r}_e$ is the equatorial radius of the star, which we have converted into Schwarzschild-type coordinates $r$ according to the equation $r=\bar{r}\exp\left[\left(\bar{\gamma }-\bar{\rho }\right)/2\right]$. To obtain the radius of the marginally (or innermost) stable  circular orbits $r_{ms}$ we have used a truncated form of the analytical approximation given as (\cite{ShSa98}),
\begin{eqnarray}
\frac{r_{ms}}{6M}&\approx &\left(1-0.54433q-0.22619q^2+0.17989Q_2-0.23002q^2+\right.\nonumber\\
&& \left. 0.26296qQ_2-0.29693q^4+0.44546q^2Q_2\right),\
\end{eqnarray}
where $q=J/M^2$ and $Q_2=-M_2/M^3$, respectively, and where $J$ is the spin angular momentum, and $M_2$ is the quadrupole moment.
Once the metric outside the rotating stars is known, from Eqs.~(\ref{tildeL})-(\ref{Omega}) we obtain the angular velocity, the specific energy and the specific angular momentum of the orbiting plasma particles in the disk. Then we calculate the integral given by Eq.~(\ref{F}), measuring the photon flux emitted by the disk surface in thermodynamical equilibrium. From the flux we obtain the temperature distribution of the disk, as well as the spectra of the emitted radiation.  For all our calculations we use an accretion rate of $\dot M=1\times10^{-12} M_{\odot}$/yr, and assume that $\cos \gamma =1$. In order to compare the thermal and emission properties of the thin accretion disks onto neutron and quark stars, we consider several classes of models in which both the neutron and the quark stars have some common characteristics.

\subsection{Accretion disks onto neutron and quark stars with same mass and angular velocity}\label{41}

 As a first example of the electromagnetic radiation characteristics of the accretion disks we consider the properties of disks around rapidly rotating neutron and quark stars having the same gravitational mass $M$ and angular velocity $\Omega $. If $M$ and $\Omega $ as well as the electromagnetic characteristics of the radiation emission from the disk could be determined by using astronomical and astrophysical observations, one may attempt to identify the type of the central star by analyzing the differences in the physical parameters (flux, temperature and spectrum) of the disk. The physical parameters of the accretion disks are obtained for a fixed total mass and  rotational frequency for all the equations of state discussed in Section~\ref{eos}.

 In Table~\ref{table1} we present the main physical parameters of the considered neutron and quark stars, with the total mass set to values around $1.8 M_{\odot}$, and the angular velocity fixed to values around $5\times 10^3 \;{\rm s}^{-1}$. In this, and the following Tables, $\rho _c$ is the central density, $M$ is the gravitational mass, $M_0$ is the rest mass, $R_e$ is the circumferential radius at
the equator, $\Omega $ is the angular velocity, $\Omega _p$ is the angular velocity of a particle in circular orbit at the equator, $T/W$ is the rotational-gravitational energy ratio, $cJ/GM_{\odot}^2$ is the angular momentum, $I$ is the moment of inertia, $\Phi_2$ gives the mass quadrupole moment $M_2$ so that $M_2=c^4\Phi_2/(G^2M^3M^3_{\odot})$, $h_+$ is the height from the surface of the last stable co-rotating circular orbit in the equatorial plane, $h_{-}$ is the height from surface of the last stable counter-rotating circular orbit in the equatorial plane, $\omega _c/\Omega$ is the ratio of the central value of the potential $\omega $ to $\Omega $, $r_e$ is the coordinate equatorial radius, and $r_p/r_e$ is the axes ratio (polar to equatorial), respectively.
\begin{table*}
\centering
\begin{tabular}{|l|l|l|l|l|l|l|l|l|l|l|l|}
\hline
EOS & DH  & RMF soft & RMF stiff & STOS 0 & STOS 0.5 & STOS 1 & BBBAV14 & BBBParis & APR & Q & CFL150 \\
\hline
$\rho_c\;[10^{15}{\rm g}/{\rm cm}^{3}]$ & 1.29 & 2.00 & 0.57 & 0.369 & 0.383 & 0.40 & 2.15 & 1.70 & 1.225 &  0.931 & 0.71  \\
\hline
$M\;[M_{\odot}]$ & 1.81 & 1.52 & 1.80 & 1.85 & 1.80 & 1.79 & 1.80 & 1.80 & 1.80 & 1.79 & 1.80\\
\hline
$M_0\; [M_{\odot}]$ & 2.05 & 1.70 & 2.00 & 2.01 & 1.95 & 1.93 & 2.08 & 2.07 & 2.11 &  2.09 & 2.09\\
\hline
$R_e [{\rm km}]$ & 12.01 & 11.34 & 15.79 & 21.03 & 22.84 & 22.72 & 10.57 & 10.98 & 10.99 &  11.79 & 12.36 \\
\hline
$\Omega [10^3{\rm s}^{-1}]$ & 4.99 & 5.00 & 5.00 & 4.90 & 4.71 & 4.45 & 5.01 & 5.00 & 5.00 &  4.79 & 5.00\\
\hline
$\Omega_p [10^3{\rm s}^{-1}]$ & 11.16 & 11.67 & 7.97 & 5.37 & 4.56 & 4.54 & 13.96 & 13.19 & 13.23 &  11.97 & 11.28\\
\hline
$T/W [10^{-2}]$ & 3.43 & 3.31 & 8.93 & 14.91 & 12.25 & 9.83 & 2.35 & 2.66 & 3.41 & 4.45 & 5.64 \\
\hline
$cJ/GM_{\odot}^2$ & 1.15 & 0.82 & 2.00 & 2.95 & 2.49 & 2.15 & 0.95 & 1.00 & 1.12 & 1.28 & 1.50 \\
\hline
$I [10^{45}{\rm g}\;{\rm cm}^2]$ & 2.03 & 1.45 & 3.52 & 5.30 & 4.65 & 4.25 & 1.66 & 1.76 & 1.98 & 2.35 & 2.63 \\
\hline
$\Phi_2 [10^{43}{\rm g}\;{\rm cm}^2]$ & 8.54 & 6.73 & 43.82 & 106.53 & 80.44 & 61.22 & 4.42 & 5.43 & 7.87 & 1.30 & 1.91\\
\hline
$h_+ [{\rm km}]$ & 6.85 & 0.00 & 0.00 & 0.00 & 0.00 & 0.00 & 3.19 & 2.69 & 0.00 &  0.00 & 0.00 \\
\hline
$h_- [{\rm km}]$ &7.48 & -2.35 & -3.40 & 7.91 & 3.87 & 2.27 & 8.14 & 7.90 & -2.11 &  0.00 & 0.00 \\
\hline
$\omega _c/\Omega [10^{-1}]$ & 5.85 & 5.52 & 4.52 & 4.10 & 4.07 & 4.08 & 6.67 & 6.34 & 5.86 &  5.27 & 5.00 \\
\hline
$r_e  [{\rm km}]$ & 9.10 & 8.90 & 12.85 & 18.00 & 19.97 & 19.91 & 7.64 & 8.06 & 8.05 &  8.86 & 9.40\\
\hline
$r_p/r_e$ & 0.88 & 0.88 & 0.72 & 0.54 & 0.53 & 0.58 & 0.92 & 0.91 & 0.90 & 0.87 & 0.84 \\
\hline
\end{tabular}
\caption{Physical parameters of the compact stars with total mass $M\approx 1.8 M_{\odot}$ and rotational frequency $\Omega\approx 5\times 10^3 \;{\rm s}^{-1}$.}
\label{table1}
\end{table*}

In Fig.~\ref{fig2} we present the total flux $F(r)$ radiated by the accretion disk around the stellar models DH, RMF soft/stiff, STOS $T=0.0$, $0.5$ and $1.0$, BBBAV14, BBBParis, APR,  Q and  CFL150, respectively, all the stars having the mass of $1.8M_{\odot }$ and the angular velocity  $5\times 10^3 \;{\rm s}^{-1}$. We can distinguish two types of behavior for the curves in this figure.

The first group contains the models with relatively small stellar radii, of the order of $10-12$ km. The stars with EOS APR, BBBAV14, BBBParis and DH belong to this group, and the inner edges of the accretion disk (indicated by the left boundary of the radial flux profile) are located at low dimensionless radii, close to each other, with $r/M\simeq 5.2-5.6$. The shape of their flux profiles is rather similar, and only the stellar model APR produces a maximal flux value higher by a factor of $1.4$ with respect to the other ones. The maximal flux values for the stars with the DH and BBBParis type EOS's are close to each other, and the model BBBAV14 has the smallest maximal value. For the models BBBAV14, BBBParis, and DH, respectively, the disk radiates its maximal flux at higher radii as compared to the APR EOS. The model RMF soft and the quark star model Q still belong to this first group, and they have accretion disks with the inner edge at $r/M\simeq5.4$. Nevertheless, the flux profiles of their disk radiation are different. The highest values for $F(r)$ among all the models of the first group are obtained for the RMF soft model. The flux emitted by the accretion disk around the quark star described by the bag model equation of state Q is lower by a factor of around $0.7$, as compared to the fluxes emitted by the disks around the neutron stars, but the flux profile is very similar to the DH model profile. The only difference between them is in the location of the inner edge of the disk. The quark star model CFL150 also belongs to the first group. The inner disk edge is located at $r/M\simeq5.6$ for this type of EOS,  but the maximal flux emitted by the disk is close to the one emitted by the BBBParis model.

The RMF stiff and the STOS type neutron stars form a second group, since the accretion disks around them emit a flux that is much smaller, as compared to the fluxes of the first group. The stellar radii of these stars are much greater than those of the stars in the first group, and their marginally stable orbits are located at even higher radii, between $r/M\simeq7$ and $11.7$, respectively. In the STOS type models, the maximum values of their fluxes are increasing, and the radii of the left boundary of $F(r)$ are decreasing, with increasing star temperatures. The EOS RMF stiff, with the smallest radius of the inner edge, gives the highest maximal flux in the second group. The temperature distribution in the disk, shown in Fig.~\ref{fig3}, follows the same pattern as the one described for the flux profiles.

\begin{figure}[tbp]
\includegraphics[width=8.15cm]{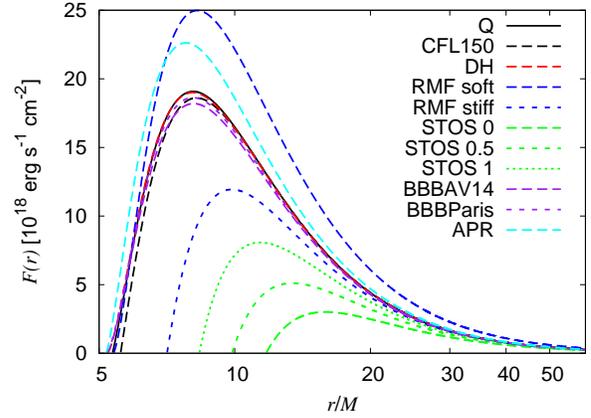}
\centering
\caption{The time-averaged flux radiated by a thin accretion disk around rotating neutron and quark stars, with the same total mass $M=1.8M_{\odot}$, and rotational velocity $\Omega =5\times10^3\;{\rm s}^{-1}$.}
\label{fig2}
\end{figure}

In Fig.~\ref{fig4} we present emission spectra of the disks, which similarly to the flux and temperature curves, also exhibit some distinctive features. Both the amplitude and the cut-off frequency of the spectra are determined roughly by the maximum values of $F(r)$: for higher maximal flux values of an accretion disk around a given stellar model we obtain higher maximal amplitudes of the disk spectra, and they are shifted to higher frequencies. This is clearly seen in Fig.~\ref{fig4} for the neutron stars with the RMF stiff  and the STOS type EOS's. However, there are some exceptions. The spectral maximum is lower for the model RMF soft as compared to the APR type EOS, but its cut-off is still located at a higher frequency. The spectral maxima are almost the same for the BBBAV14 and BBBParis models, but the cut-off frequency is higher for the BBBAV14 type EOS. The spectral maxima for all the models in the plot are located between $2\times 10^{16}$ and $4\times 10^{16}$ Hz.
The spectral features of the neutron star model DH and of both types of quark star models are essentially the same, and thus only the morphological properties of the accretion disk, i.e., the radial distribution of the flux or the disk temperature, can be used to discriminate these type of compact stellar objects from each other.

\begin{figure}[tbp]
\centering
\includegraphics[width=8.15cm]{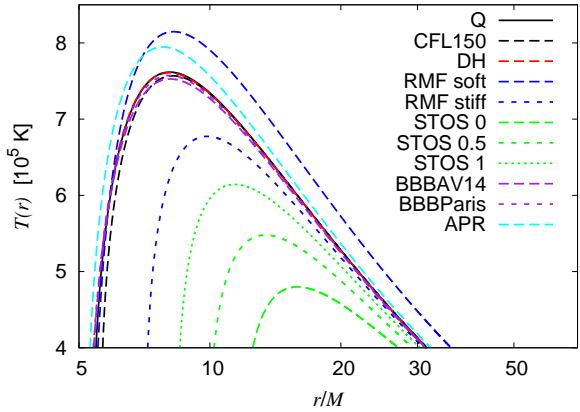}
\caption{The temperature distribution in a thin accretion disk around rotating neutron and quark stars with the same total mass $M=1.8M_{\odot}$, and rotational velocity $\Omega =5\times10^3\;{\rm s}^{-1}$.}
\label{fig3}
\end{figure}

\begin{figure}[tbp]
\centering
\includegraphics[width=8.15cm]{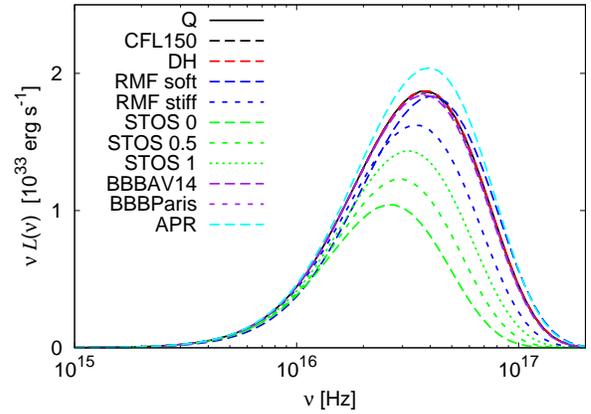}
\caption{The emission spectrum of the accretion disk around rotating neutron and quark stars with the same total mass $M=1.8M_{\odot}$, and rotational velocity $\Omega =5\times10^3\;{\rm s}^{-1}$.}
\label{fig4}
\end{figure}

\subsection{Accretion disks onto neutron and quark stars rotating with Keplerian velocities}

As a second case of comparison between the neutron and quark star properties we consider the Keplerian models of the neutron and quark stars,  where they rotate at the maximal frequency allowed by their physical properties. We fix the central density for all the stellar models to $\rho _c=10^{15}\;{\rm g}/{\rm cm}^{3}$, and we compute the physical parameters of all the stellar models for the case of the Keplerian rotation. The results are represented in Table~\ref{table2}.

\begin{table*}
\begin{center}
\begin{tabular}{|l|l|l|l|l|l|l|l|l|l|l|l|}
\hline
EOS & DH  & RMF soft & RMF stiff & STOS 0 & STOS 0.5 & STOS 1 & BBBAV14 & BBBParis & APR & Q & CFL300 \\
\hline
$\rho_c\;[10^{15}{\rm g}/{\rm cm}^{3}]$ & 1.00  & 1.00 & 1.00 & 1.00 & 1.00  & 1.00 & 1.00 & 1.00 & 1.00 & 1.00 & 1.00\\
\hline
$M\;[M_{\odot}]$ & 1.78 & 1.84 & 2.78 & 3.49 & 3.42 & 3.35 & 1.53 & 1.54 & 2.10 & 2.79 & 4.76\\
\hline
$M_0\; [M_{\odot}]$ & 1.98 & 2.03 & 3.23 & 4.15 & 4.03 & 3.92 & 1.68 & 1.70 & 2.40 & 3.29 & 5.98\\
\hline
$R_e [{\rm km}]$ & 16.24 & 17.78 & 18.42 & 19.15 & 19.24 & 19.43 & 15.58 & 15.74 & 15.52 & 17.07 & 20.50 \\
\hline
$\Omega [10^3{\rm s}^{-1}]$ & 7.44 & 6.91 & 7.74 & 7.99 & 7.81 & 7.60 & 7.37 & 7.29 & 8.96 &  9.07 & 8.59\\
\hline
$\Omega_p [10^4{\rm s}^{-1}]$ & 7.44 & 6.91 & 7.74 & 7.99 & 7.81 & 7.60 & 7.37 & 7.29 & 8.97 & 9.08 & 8.79\\
\hline
$T/W [10^{-1}]$ & 1.14 & 1.62 & 1.54 & 1.48 & 1.38 & 1.27 & 1.12 & 1.12 & 1.95 & 2.27 & 2.18 \\
\hline
$cJ/GM_{\odot}^2$ & 2.15 & 2.91 & 5.94 & 8.98 & 8.31 & 7.66 & 1.61 & 1.64 & 3.83 &  7.15 & 1.96 \\
\hline
$I [10^{45}{\rm g}\;{\rm cm}^2]$ & 2.54 & 3.70 & 6.74 & 9.88 & 9.34 & 8.86 & 1.92 & 1.97 & 3.75 & 6.91 & 2.01 \\
\hline
$\Phi_2 [10^{44}{\rm g}\;{\rm cm}^2]$ & 34.02 & 73.57 & 99.54 & 124.77 & 110.35 & 97.48 & 27.12 & 27.84 & 69.85  & 13.15 & 33.29 \\
\hline
$h_+ [{\rm km}]$ & 0.00 & 0.00 & -3.59 & 1.65 & 1.37 & 1.06 & 0.00 & 0.00 & -3.14 &  -3.14 & -2.83 \\
\hline
$h_- [{\rm km}]$ & 6.94558 & -3.91 & 17.75 & 24.54 & 23.11 & 21.64 & 4.89 & 4.91 & 13.63 &  0.13 & -0.08 \\
\hline
$\omega _c/\Omega [10^{-1}]$ & 5.44 & 5.15 & 6.90 & 7.95 & 7.88 & 7.82 & 4.94 & 4.95 & 6.07 &  6.96 &  8.82 \\
\hline
$r_e  [{\rm km}]$ & 13.37 & 14.72 & 13.66 & 13.05 & 13.35 & 13.72 & 13.13 & 13.27 & 11.91 &  11.99 & 11.06\\
\hline
$r_p/r_e$ & 0.55 & 0.49 & 0.52 & 0.54 & 0.55 & 0.56 & 0.55 & 0.55 & 0.47 & 0.44 & 0.51 \\
\hline
\end{tabular}
\end{center}
\caption{Physical parameters of the neutron and quark stars rotating at Keplerian frequencies. All stars have the same central density $\rho _c$.}
\label{table2}
\end{table*}

The flux curves for these configurations are shown in Fig.~\ref{fig5}.
Here each EOS type has very distinctive features. Only the STOS type stars, and the models BBBAV14 and BBBParis produce similar flux profiles.  Nevertheless, there is a temperature dependence for the models STOS
in the location of the inner disk edge ($r/M=4.1-4.2$), and of the maximal value of the photon flux emerging from the disk: with increasing temperatures we obtain greater radii for the inner edge of the disk, and the maximal flux values are inversely proportional to the temperature of the star. For the models BBBAV14 and BBBParis, we see that the location of the inner edge of the accretion disks around these types of neutron stars are essentially the same ($r/M\simeq7.1$), but the model BBBAV14 produces a higher maximal flux than the BBBPAris type EOS does. Although the thin disk rotating around the APR type star has an inner edge at only a little bit lower radius than the disks for the models BBBAV14 and BBBParis, respectively, its maximal radiated away flux is a factor of $1.4$ smaller than those for the neutron stars with the latter types of EOS.

The inner edge of the accretion disk around the quark star CFL300 is located at the lowest radius ($r/M\simeq3.6$), as compared with the other EOS models in this configuration, whereas the model RMF soft has the greatest radius for the inner edge ($r/M\simeq9.5$). The flux values for the RMF soft type star are the smallest ones as well. The maximal flux for the model DH is the greatest in this group, and it is at least a factor of 3.5 greater than the one for the model RMS stiff.
The flux profiles of the Q type quark star and of the neutron star with the RMF stiff type EOS are very similar, but the inner edge of the disk rotating around the neutron star is located at a lower radius ($r/M\simeq5.5$ for the quark star, and $r/M\simeq 5.1$ for the neutron star). For the radial distributions of the disk temperature we obtain the same characteristic behavior and groups of models,  which we present in Fig.~\ref{fig6}.

The spectral features, as seen in Fig.~\ref{fig7}, are strongly influenced by the flux and the temperature distributions of the disk. For the case of the Keplerian rotation the emission spectra of the disk are rather different from those of the previous configuration, with same gravitational mass and angular velocity.
The proportionality between the location of the inner disk edge and the maximal amplitudes of the spectra holds for this case as well. Therefore the greatest spectral maximum is produced by the quark star model CFL300, whereas the smallest one is obtained for the model RMF stiff.
The STOS models have also high amplitudes in their spectra, and their maximal values, in contrast with the previous case, are inversely proportional to the temperature.

There are no big differences between the cut-off frequencies in the spectra, which are around $2\times10^{16}-3\times10^{16}$Hz. The spectrum belonging to the RMF stiff type EOS has its maximum at the lowest frequency, and the spectral maxima for the models DH, BBBAV14 and BBBParis are located at the highest frequencies. Since the spectra for the latter two models are very similar, we can use only the differences in the flux maxima to discriminate between them. This is also the case for the quark star model Q and the RMF stiff type neutron star as well, where the only discriminating factor is the difference in the locations of the inner disk edges. The models APR and DH have also similar spectra, but both their maxima and cut-off frequencies exhibit some differences.

\begin{figure}[tbp]
\centering
\includegraphics[width=8.15cm]{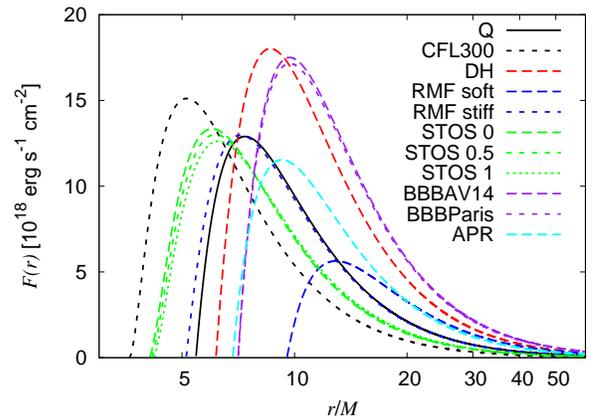}
\caption{The time-averaged flux radiated by a thin accretion disk around neutron and quark stars rotating at the Keplerian velocity. All the stars have the same central density $\rho _c=10^{15}$ g/cm$^3$.}
\label{fig5}
\end{figure}

\begin{figure}[tbp]
\centering
\includegraphics[width=8.15cm]{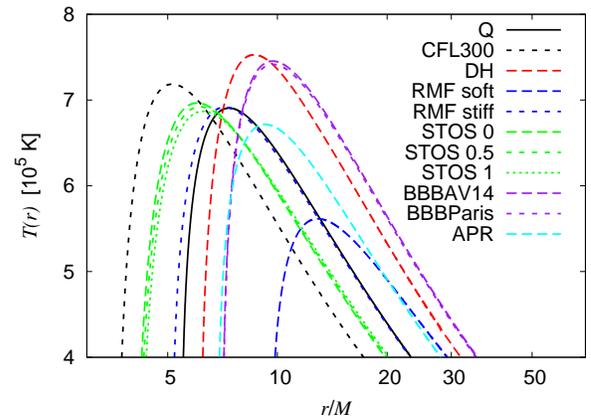}
\caption{The temperature distribution by a thin accretion disk around neutron and quark stars rotating at the Keplerian velocity. All the stars have the same central density $\rho _c=10^{15}$ g/cm$^3$.}
\label{fig6}
\end{figure}

\begin{figure}[tbp]
\centering
\includegraphics[width=8.15cm]{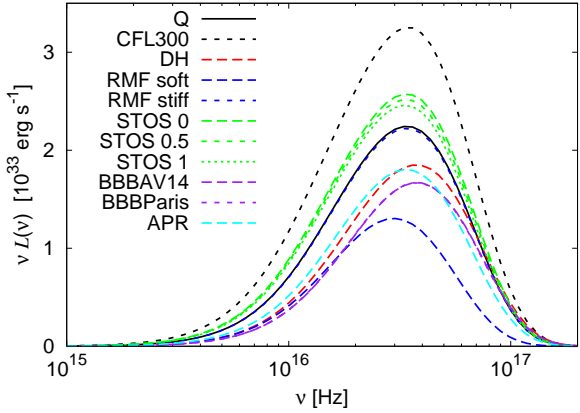}
\caption{The emission spectrum of the accretion disk around neutron and quark stars rotating at the Keplerian velocity. All the stars have the same central density $\rho _c=10^{15}$ g/cm$^3$.}
\label{fig7}
\end{figure}

\subsection{Accretion disks onto neutron and quark stars with fixed central density and eccentricity}

As a third example of the comparison between the radiation properties of the thin accretion disks onto neutron and quark stars we consider the case in which the rotating central compact objects have the same central density and eccentricity $r_p/r_e$, respectively, where $r_p$ is the polar radius and $r_e$ is the equatorial radius of the star. The basic fundamental parameters of the rotating neutron stars and quark stars are presented in Table~\ref{table3}.

\begin{table*}
\begin{center}
\begin{tabular}{|l|l|l|l|l|l|l|l|l|l|l|l|}
\hline
EOS & DH  & RMF soft & RMF stiff & STOS 0 & STOS 0.5 & STOS 1 & BBBAV14 & BBBParis & APR & Q & CFL300 \\
\hline
$\rho_c\;[10^{15}{\rm g}/{\rm cm}^{3}]$ & 1.00 & 1.00 & 1.00 & 1.00 & 1.00 & 1.00 & 1.00 & 1.00 & 1.00 & 1.00 & 1.00\\
\hline
$M\;[M_{\odot}]$ & 1.69 & 1.72 & 2.62 & 3.32 & 3.27 & 3.22 & 1.45 & 1.46 & 2.05 &  2.80 & 4.50\\
\hline
$M_0\; [M_{\odot}]$ & 1.88 & 1.91 & 3.04 & 3.95 & 3.86 & 3.77 & 1.59 & 1.60 & 2.35 &  3.30 & 5.67\\
\hline
$R_e [{\rm km}]$ & 20.01 & 19.47 & 20.95 & 22.17 & 22.66 & 23.24 & 19.32 & 19.51 & 16.14 &16.95 & 21.59 \\
\hline
$\Omega [10^3{\rm s}^{-1}]$ & 6.82 & 6.71 & 7.34 & 7.49 & 7.28 & 7.05 & 6.74 & 6.67 & 8.91 & 9.07 & 8.36\\
\hline
$\Omega_p [10^3{\rm s}^{-1}]$ & 5.26 & 5.63 & 6.06 & 6.18 & 5.93 & 5.67 & 5.15 & 5.09 & 8.19 & 9.25 & 7.50\\
\hline
$T/W [10^{-2}]$ & 9.14 & 13.77 & 12.75 & 12.26 & 11.37 & 10.45 & 8.91 & 8.89 & 18.64 &  22.87 & 19.27 \\
\hline
$cJ/GM_{\odot}^2$ & 1.72 & 2.33 & 4.75 & 7.39 & 6.90 & 6.41 & 1.28 & 1.30 & 3.58 &  7.22 & 16.55 \\
\hline
$I [10^{45}{\rm g}{\rm cm}^2]$ & 2.22 & 3.06 & 5.68 & 8.66 & 8.32 & 7.99 & 1.67 & 1.71 & 3.53 & 6.99 & 1.73 \\
\hline
$\Phi_2 [10^{44} {\rm g}{\rm cm}^2]$ & 25.00 & 53.68 & 71.23 & 92.12 & 82.80 & 74.09 & 19.86 & 20.39 & 63.51 & 13.34 & 2.55 \\
\hline
$h_+ [{\rm km}]$ & 0.00 & 0.00 & 0.00 & 0.00 & 0.00 & 0.00 & 0.00 & 0.00 & 0.00 &  -3.09 & 1.63 \\
\hline
$h_- [{\rm km}]$ & 1.28 & 5.25 & 12.07 & 18.34 & 16.85 & 15.26 & 0.00 & 0.00 & 12.15 &  -0.34 & 35.70 \\
\hline
$\omega _c/\Omega [10^{-1}]$ & 5.25 & 4.98 & 6.71 & 7.79 & 7.74 & 7.68 & 4.75 & 4.76 & 6.01 & 6.97 & 8.69 \\
\hline
$r_e  [{\rm km}]$ & 17.37 & 16.71 & 16.70 & 16.69 & 17.31 & 18.01 & 17.07 & 17.25 & 12.69 & 13.41 & 8.75\\
\hline
$r_p/r_e$ & 0.45 & 0.45 & 0.45 & 0.45 & 0.45 & 0.45 & 0.45 & 0.45 & 0.45 & 0.45 & 0.45 \\
\hline
\end{tabular}
\end{center}
\caption{Physical parameters of the rapidly rotating compact stars with same central density and eccentricity.}
\label{table3}
\end{table*}

The physical parameters of the rotating stars have been obtained by choosing  for all the considered equations of state a central density $\rho _c=1\times 10^{15}$ erg/cm$^3$, and an eccentricity $r_p/r_e=0.45$, respectively. We present the radial distribution of the energy flux $F(r)$  in this stellar configuration for the neutron star models DH, RMF soft/stiff, STOS $T=0.0$, $T=0.5$ MeV and $T=1.0$ MeV, BBBAV14, BBBParis, APR, and for the quark star models Q and CFL300, respectively, in Fig.~\ref{fig8}.

\begin{figure}[tbp]
\centering
\includegraphics[width=8.15cm]{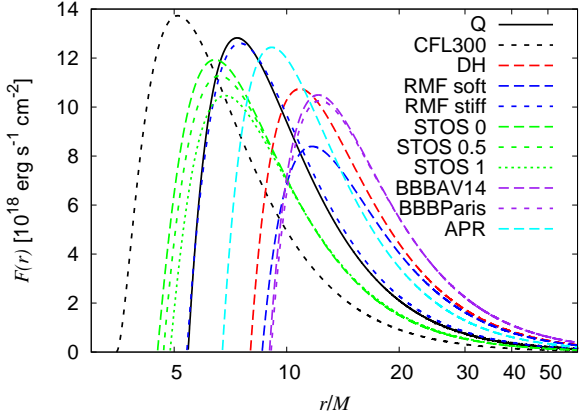}
\caption{The time-averaged flux radiated by a thin accretion disk around rotating neutron and quark stars, with the same central density ($\rho _c=10^{15}$ g/cm$^3$) and eccentricity ($r_p/r_e=0.45$).}
\label{fig8}
\end{figure}

Fig.~\ref{fig8} shows that different equations of state with the same central density and eccentricity produce rather different geometries around the compact central object, and in turn determine rather distinct emissivity properties of the accretion disks.
The distribution of the radial flux profiles is similar to the Keplerian case, shown in Fig.~\ref{fig5}, but with some significant differences.
The differences between the amplitudes of the disks radiation for the central objects with the same $\rho_c$ and $r_e/r_p$ ratio are smaller as compared to the ones obtained in Fig.~\ref{fig5}. This is due to the considerable decrease of the maximal flux values for the models DH, BBBVA14 and BBBParis, respectively, as compared to the maxima reached in the Keplerian case. The other EOS types produce more or less the same, or somewhat higher, flux maxima as compared to the case of the Keplerian rotation. By considering the inner edges of the disk, we find that the radial distribution of their location is very similar to the one seen in Fig.~\ref{fig5}.
Nonetheless, the differences between the flux profiles for the STOS models are bigger, and there is also a shift of the inner edge of the disk to higher radii for the models DH, BBBVA14 and BBBParis, respectively, which determines a significant reduction in their flux values.
These features can also be seen in Fig.~\ref{fig9}, where we display the temperature distribution of the disk surface in thermodynamical equilibrium.

\begin{figure}[tbp]
\centering
\includegraphics[width=8.15cm]{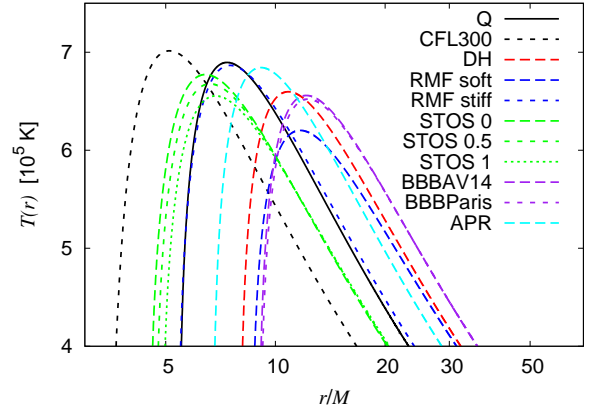}
\caption{Temperature distribution in a thin accretion disk around rotating neutron and quark stars, with the same central density ($\rho _c=10^{15}$ g/cm$^3$) and eccentricity ($r_p/r_e=0.45$).}
\label{fig9}
\end{figure}

Important differences are also present in the disk spectra. The black-body spectrum of the equilibrium thermal radiation emitted from the accretion disk for each type of star is shown in Fig.~\ref{fig10}.
The quark stars together with the STOS type neutron stars produce greater maximal amplitudes of the disk emission spectra as compared to the other neutron stars in this group. The inverse proportionality between the temperature of the star and the maximal amplitudes of the spectra holds again for the STOS models, similarly to the Keplerian case. However, the spectrum of the quark star with EOS Q and the spectrum of the neutron star with the RMF stiff type EOS are decoupled in this case, and the disk spectrum for the quark star is more similar to the spectrum of the model STOS 1. The latter has a slightly lower cut-off frequency, and these two types of EOS could be discriminated from each other by comparing the radial distributions of the flux and of the temperature of their disk. The spectra of the group consisting of the models DH, RMF soft, BBBAV14 and BBBParis are similar. In fact, the spectra coincide for the latter two models. For this case the different spatial distributions of the disk radiation can help again in the discrimination. There is no such a problem with the APR type EOS, since it produces somewhat higher spectral amplitudes as compared to the other cases. The maxima of all the spectra are located in a very narrow frequency range between $2\times10^{16}$ and $2.2\times10^{16}$Hz, but the numerical values of the flux maxima are rather different, more or less following those of the flux maxima.}

\begin{figure}[tbp]
\centering
\includegraphics[width=8.15cm]{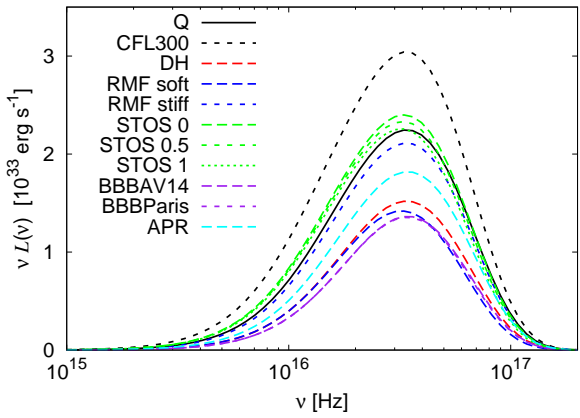}
\caption{The emission spectrum of the accretion disk around rotating neutron and quark stars, with the same central density ($\rho _c=10^{15}$ g/cm$^3$) and eccentricity ($r_p/r_e=0.45$).}
\label{fig10}
\end{figure}

\section{Discussions and final remarks}

In the present paper we have considered the basic physical properties of
matter forming a thin accretion disc around rapidly rotating neutron and quark stars. The physical parameters of the disc - effective potential, flux and emission spectrum profiles - have
been explicitly obtained for several equations of state
of the neutron matter and for two types of quark stars, respectively.  All the astrophysical quantities, related to the observable properties of the accretion disc, can be obtained from
the metric of the central compact object. Due to the differences in the space-time
structure, the quark stars  present some very important differences with respect to the disc properties, as
compared to the neutron stars. This procedure allows not only to discriminate between neutron and quark stars, but also gives some distinct signatures to differentiate between neutron stars described by different equations of state, and to distinguish between the different phases of the quark matter.

Another observable physical parameter of the disk that could help distinguishing between different classes of compact objects is the efficiency $\epsilon$ of the conversion of the accreting mass into radiation, given by Eqs.~ (\ref{epsilon}) and Eq.~(\ref{epsilon2}). The values of $\epsilon$ measure how efficient the energy generating mechanism of the mass accretion is. The amount of energy released by the matter leaving the marginally stable orbit, or the inner edge of the disk touching the surface of the star, and being transferred to the star, is the binding energy $\widetilde{E}_{ms}$ or $\widetilde{E}_{e}$. We present the efficiency for the energy conversion for the different equations of state considered in the present paper in Table~\ref{table4}. The first two lines of Table~\ref{table4} contain the conversion efficiency of the rotating compact stars with the same total mass and angular velocity.

For this case the values of $\epsilon$ are in the range 4\% and 8\%, indicating that in this case there is no significant difference in the efficiency of the mass-to-radiation conversation mechanism. The less efficient conversion mechanism is provided by the STOS models, especially at the zero temperature of the star, whereas the APR type EOS provides the most efficient conversion mechanism with $\epsilon =7.4\%$. The quark stars with $\epsilon=6.8$\% are also efficient engines for the conversion of the accreted mass into outgoing radiation.

The values of $\epsilon $ for the compact general relativistic objects in Keplerian rotation are given in the second two lines of Table~\ref{table4}. The case of stars with the same total mass and angular velocity is represented in (a), stars in Keplerian rotation are described in (b), while the case of the stars with the same central density and eccentricity is shown in (c), respectively.

\begin{table*}
\begin{center}
\begin{tabular}{|l|l|l|l|l|l|l|l|l|l|l|l|}
\hline
EOS & DH  & RMF soft & RMF stiff & STOS 0 & STOS 0.5 & STOS 1 & BBBAV14 & BBBParis & APR & Q & CFL300/150 \\
\hline
(a)  $r_{in}$[km]  &  13.88 & 12.10 & 18.81 & 32.18 & 26.33 & 22.11 & 13.83 & 13.76 & 13.85 & 14.24 & 14.84 \\
     $\;\;\;\;\;\;\;\;\epsilon $    & 0.0679 & 0.0667 & 0.0589 & 0.0400 & 0.0463 & 0.0527 & 0.0673 & 0.0682 & 0.0742 & 0.0685 & 0.0683 \\
\hline
(b) $r_{in}$[km]  & 16.24 & 25.97 & 21.13 & 21.17 & 21.00 & 20.87 & 16.02 & 16.17 & 21.22 & 22.51 & 23.83 \\
     $\;\;\;\;\;\;\;\;\epsilon $   & 0.0676 &  0.0488 & 0.0802 & 0.0929 & 0.0913 &  0.0894 & 0.0611 & 0.0610 & 0.0663 & 0.0814 & 0.1172 \\
\hline
(c) $r_{in}$[km] & 20.01 & 21.95 & 20.96 & 22.18 & 22.66 & 23.24 & 19.32 & 19.52 & 20.45 & 22.61 & 23.26 \\
    $\;\;\;\;\;\;\;\;\epsilon $   &0.0557 & 0.0528 &0.0768 & 0.0869 & 0.0845 & 0.0819 & 0.0503 & 0.0502 & 0.0666 & 0.0814 & 0.1097\\
\hline
\end{tabular}
\end{center}
\caption{The radius of the inner disk edge $r_{in}$ and the efficiency $\epsilon$.}
\label{table4}
\end{table*}

 In this case the quark star model CFL300 provides the most efficient mechanism for the conversion, with $\epsilon=11.7$\%. The quark stars described by the Q EOS have also a high value of the conversion efficiency, with $\epsilon=8.15$\%. The conversion efficiency for the models STOS are much higher than the ones in the previous stellar configuration. Their efficiencies can reach for this case a value of $\epsilon =9\%$. The other models have efficiencies of the order of 5-6\%.

The last two lines in Table~\ref{table4} contains the values of the efficiency for stellar models with the same eccentricity and central density, respectively.
The range of the values of $\epsilon $ is similar to the previous case, although in general the numerical values are somewhat lower. We obtain again the highest efficiency for the model CFL300, and the smallest ones for the neutron stars DH, RMF soft, BBBAV14 and BBBParis, respectively, with $\epsilon $ about 5\%. The STOS type neutron stars and the Q type quark stars are more efficient, with  $\epsilon =7-8\%$.

As shown by the flux integral in Eq.~(\ref{F}), and the explicit expressions of the specific energy, specific angular momentum and angular velocity given by Eqs.~(\ref{tildeE}), (\ref{tildeL}) and (\ref{Omega}),  the rather different characteristics of the radial flux distribution over the accretion disk, the disk spectra and the conversion efficiency are due to the differences between the metric potentials of the neutron and of the quark stars, respectively. Even if the total mass and the angular velocity are the same for each type of the rotating central object, producing similar values of $\Omega$, $\widetilde E$ and $\widetilde L$, the radiation properties of the accretion disks around these objects exhibit discernible differences. The reason is that the proper volume, and in turn the function $\sqrt{-g}$, used in the calculation of the flux integral, are highly dependent on the behavior of the metric component $g_{rr}=(\partial \overline{r}/\partial r)^2 g_{\overline{r}\overline{r}}$. The latter contains the $r$-derivatives of the metric functions $\rho(r)$ and $\gamma(r)$, via the coordinate transformation between $\overline{r}$ and $r$, which are extremely sensitive to the slope of $\rho(r)$ and $\gamma(r)$. As a result, though the inner edges of the disks are located at almost the same radii, the amplitudes of the energy fluxes emerging form the disk surface and propagating in any solid angle may have considerable differences for different equations of state. These features also give the distinctive features in the disk spectra for the various types of central stars.

In conclusion, the observational study of the thin accretion disks around rapidly rotating compact objects can provide a powerful tool in distinguishing between standard neutron stars and stars with exotic equations of state, that have underwent, for example, a phase transition from the neutron phase to a quark phase, as well as for discriminating between the different equations of state of the dense matter.

\section*{Acknowledgments}

 We would like to thank to the anonymous referee for suggestions and comments that helped us to significantly improve the manuscript. Z. K. is indebted to the colleagues in the Department of Physics and Center for Theoretical and Computational Physics of the University of Hong Kong for their support and warm hospitality. K. S. C. is supported by the GRF grant number HKU 7013/06P of the government of the Hong Kong SAR. The work of T. H. is supported by the GRF grant number HKU 702507 of the Government of the Hong Kong SAR.

\end{document}